\documentclass[twocolumns]{aa}
\usepackage{graphicx}
\usepackage{comment}
\usepackage{url}
\usepackage[breaklinks,colorlinks,citecolor=blue]{hyperref}
\usepackage{caption}
\usepackage{subcaption}
\usepackage{txfonts}
\newcommand{\vR}{v_{\rm R}}
\newcommand{\sigmaR}{\sigma_{\rm R}}
\newcommand{\mean}[1]{\langle #1 \rangle}
\newcommand{\ratioR}{\mean{\vR}/\sigmaR}

\begin{document}

   \title{Right round: onset and long-term evolution of rotation in star clusters}

\author{E. Dalessandro\inst{1}\thanks{\email{emanuele.dalessandro@inaf.it}} 
\and    A. Della Croce \inst{2,1} \and 
    E. Vesperini \inst{2} \and
    M. Cadelano \inst{3,1} \and
    S. Leanza \inst{1} \and
    G. Ettorre \inst{1,3} \and
    M. Hughes \inst{2}   
}

\institute{
INAF -- Astrophysics and Space Science Observatory of Bologna, via Gobetti 93/3, I-40129 Bologna, Italy
    \and 
Department of Astronomy, Indiana University, Swain West, 727 E. 3rd Street, IN 47405 Bloomington, USA
    \and 
Department of Physics and Astronomy Augusto Righi, University of Bologna, via Gobetti 93/2, I-40129 Bologna, Italy
}

   \date{Received XX; accepted XXX XX, XXXX}
 
  \abstract{We present the results of a detailed kinematic analysis of a significant fraction of the known population of Galactic 
  star clusters
  aimed at constraining the physical mechanisms driving the onset and evolution of cluster rotation. 
  Our study reveals for the very first time the presence of rotation in clusters at any age, with about 
  $25\%-30\%$ of systems in the sample showing significant 
  evidence of rotation. This result increases by a factor of $\sim5$ the number of clusters identified as rotators so far and it finally enables an observational reading of cluster rotation as a function of time.     
  Young ($<500$ Myr) clusters show a larger range of rotation velocities than older systems. In addition, at young ages we observe a significantly larger fraction ($50\%-60\%$) of rotating systems than at older ones ($\sim 15\%$).
  These purely empirical results are compatible with rotation being imprinted during the very early stages of cluster formation and early evolution and then being progressively erased
  by the long-term effects of dynamical evolution.
  For the sub-sample of clusters for which we were able to perform a full 3D analysis, we calculated the angle between the internal rotation axis and that of the cluster orbital motion.
 Interestingly, while for clusters with an age smaller than their orbital period we observe similar fractions of prograde and retrograde systems, more evolved clusters appear to be preferentially prograde. We argue that such a behavior is in qualitative agreement with the expectations for the evolution of systems in which primordial rotation was imprinted by the parent molecular cloud and/or by the following hierarchical cluster assembly processes, and in which internal cluster dynamics and interactions with the Galactic field have induced a torque-driven alignment between cluster rotation and orbital motion.}

   \keywords{astrometry - stars: formation stars: kinematics and dynamics open clusters and associations: general - galaxies: star clusters: general - globular clusters: general}

\titlerunning{Stellar cluster rotation}
\authorrunning{Dalessandro et al.}
   \maketitle



\section{Introduction}
The study of star cluster formation and evolution has witnessed a renewed and significant interest in recent years both from 
the theoretical and observational point of view (e.g., \citealt{pascale_m_25,lahen25,calura19,pascale_r_25,taylor25,livernois21,dalessandro21a,dalessandro24,dellacroce23}).
Particular attention has been paid to the very early stages of evolution, when clusters are expected to undergo gas expulsion, 
violent relaxation and to be possibly evolving through hierarchical mass assembly and toward a virial equilibrium state (e.g., \citealt{livernois21,dalessandro21a,dellacroce23,farias23,polak24}).

The environment in which stars form determines 
a number of key properties that characterize the stellar clusters we observe today, such as their initial
mass function, structural properties, stellar binarity/multiplicity and the star-by-star chemical abundance differences routinely observed in massive stellar 
clusters (e.g., \citealt{bastianlardo18,gratton19}). In addition, the dynamics of the contracting embedded
molecular cloud cores is expected to be imprinted on the emerging stellar populations, and in particular on the surviving star cluster. As a consequence, internal cluster kinematics turns out to be a key tool to shed new light on the physical processes involved in cluster formation and evolution: the motion of stars in nascent star clusters is inherited from the parent gas cloud, probing the initial conditions of cluster formation. 
The combination of Gaia \citep{gaiaDR3}, and in particular its third data release (Gaia DR3), and wide spectroscopic surveys such as Gaia-ESO, APOGEE or GALAH for example, have finally provided an unprecedented multidimensional view of stellar clusters possibly securing 3D positions and velocities, and detailed photometry.
Indeed, several works have come out in recent years about the number and properties of star clusters and nearby star-forming regions (e.g., \citealt{cantat18,cantatgaudin_etal2020,pang21,hunt23,hunt24}) as well as about their individual structural and kinematic properties (e.g., \citealt{kuhn19,dalessandro21a,beccari20,jerabkova19,swiggum21,dellacroce23,dellacroce24,jadhav24,guilherme23}).

In this framework, in a recent paper \citep{dellacroce24} we focused on the early expansion and survival of young star clusters providing the first 
empirical description of the characteristic timescale during which cluster expansion 
plays a dominant role in the cluster dynamics.
In particular, we found that a remarkable fraction (up to $80\%$) of clusters younger than $\sim30$ Myr is
currently experiencing significant expansion, while older systems are mostly compatible with an equilibrium configuration. We showed that such an observed pattern is in general qualitative agreement with what 
expected for systems undergoing violent relaxation and possibly evolving toward a final virial equilibrium state. 
Additional processes associated with gas expulsion and mass loss due to stellar evolution likely play a role in driving the observed expansion. 

In this work we focus on cluster rotation as a complementary key kinematic property of star clusters.
There is general consensus about the fact that star cluster rotation plays a critical role in their long-term dynamical evolution. For example, rotation can speed up 
the collapse of the core through the gravo-gyro instability or it can significantly increase the escape 
rate for clusters in a strong tidal field (e.g., \citealt{einsel99,kim02,ernst07}).
However, the origin of star clusters' rotation is still a matter of investigation. 
A number of theoretical studies have shown that many physical factors can play an important role in this context. 
Angular momentum could be either inherited from the parent molecular cloud or imprinted by gas accretion in the formation phase. In addition, dynamical interactions among sub-clumps, the torque of the host galaxy gravitational potential, and stellar feedback on the left-over gas can contribute to shape clusters' rotational properties \citep{vesperini14,mapelli17,tiongco17,ballone21,chen21,lahen20,lahen25}.

\begin{figure*}
    \centering
    \includegraphics[width=0.95\linewidth]{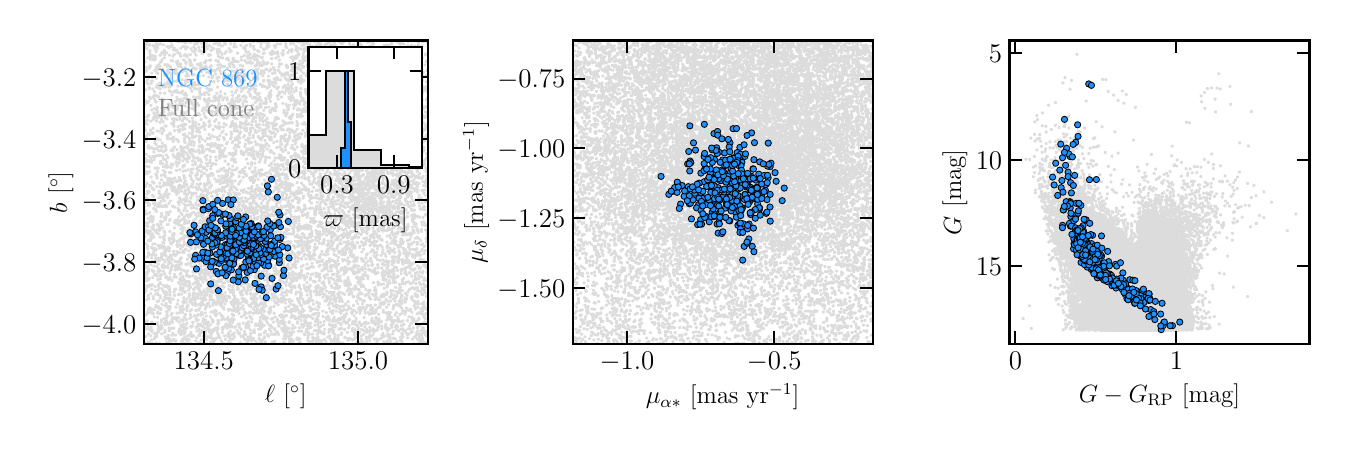}
    \caption{NGC~869 member properties: 
    the left panel shows the spatial distribution in Galactic coordinates, in the middle panel the vector-point diagram is presented while the right panel displays the distribution in the color-magnitude diagram. 
    The small inset shows the parallax distribution.
    Gray points are the starting sample of {\it Gaia} sources used in the clustering analysis, while in blue are NGC~869 member stars.}
    \label{fig:clustering_ngc869}
\end{figure*}

Until very recently, the main observational information about cluster rotation was mostly based on old Galactic 
globular clusters studies (e.g.; \citealt{anderson03,glen06,bellazzini12,fabricius14,bellini17,kamann18,ferraro18,lanzoni18a,bianchini18,sollima19,vasiliev21,dalessandro21b,dalessandro24}). Indeed, thanks to the use of high-multiplex multi-object 
and integral field spectrographs, and more recently, thanks to the Gaia astrometric mission \citep{gaiaDR3},  
these analysis have demonstrated that
the vast majority of GCs shows signature of internal rotation with amplitudes that can be as large as half of their internal velocity dispersion.
According to a number of numerical studies (see e.g., \citealt{einsel99,ernst07,tiongco17}),
present-day signatures of rotation could be the relic of a stronger internal rotation set at the epoch of 
the clusters' formation
(e.g., \citealt{vesperini14,mapelli17})
and gradually altered and erased as a 
result of the effects of angular momentum transfer and loss due to internal dynamical processes and star escape.
Indeed, observational studies have convincingly shown that the strength of rotation decreases as a function of GCs' dynamical
age \citep{kamann18,bianchini18,sollima19,dalessandro24}.

Because of the poor statistics and of the presence of pretty complex and possibly out-of-equilibrium morphological and
kinematic patterns (such as contraction, expansion and tidal tails), similar investigations have been proved to be
extremely challenging for younger and less massive clusters. These limitations have been 
progressively overcome thanks mainly to Gaia, which has significantly increased the availability of accurate proper 
motions for large samples of stars in star clusters. Recent extensive works (e.g., \citealt{kuhn19,guilherme23,jadhav24}) have provided first attempts to systematically study cluster rotation among Galactic open clusters (OCs). 
However, mainly due to the different data and methods adopted by the different authors, as of now it appears to be extremely hard to agree on the identification of candidate rotating clusters. In fact, while recent analysis identify as possible rotating clusters a few tens of systems in total, they 
agree only on the rotation of three of them, namely Stock~2, IC~2602 and Ruprecht~147, 
thus indicating the need for further work to identify new and more robust methods to measure cluster rotation and follow its evolution in statistically
significant samples of clusters.

To move a step forward in our understanding of the origin and evolution of cluster rotation, in this work we perform a 
systematic and homogeneous study based on state-of-the-art observations along the 
three velocity components and on a data analysis approach able to deal with 
relatively small statistical samples and to account for the morphological and kinematic complexities commonly observed in young OCs \citep{dalessandro24}. 
The paper is organized as follows. In Sect.~\ref{sec:data} we present the observational dataset and the cluster sample. In Sect.~\ref{sec:kin_analysis} we detail on the adopted kinematic analysis. In Sect.~\ref{sec:results} we present the main results and in Section~\ref{sec:lit} we compare them with those available in the literature.
In Sect.~\ref{sec:prograde} we report the results of the full 3D analysis and we discuss the possible link between cluster rotation and their orbit within the Galaxy. In Sect.~\ref{sec:summary} we summarize and discuss our main findings and their possible implications.

\section{Data and selected sample} \label{sec:data}

We adopted the list of 2017 clusters identified by \citet{cantatgaudin_etal2020} using Gaia DR2 as the starting sample for our analysis. 
Then to take full advantage of the most recent and accurate Gaia DR3 data release for the definition of cluster member stars,
we performed an independent membership analysis 
in the five-dimensional space of Galactic coordinates, proper motions and parallaxes 
($l$, $b$, $\mu_{\alpha^{*}}$, 
$\mu_{\delta}$ and $\varpi$) by extending the approach already adopted by \citet{dellacroce24}.
Briefly, for each cluster in the \citet{cantatgaudin_etal2020} catalog, we estimated the parameters $R_{\rm search, sky}$ and $R_{\rm search, PM}$ as the radii enclosing the 95\% of cluster members in sky coordinates and PM space. Then, from the Gaia Archive we queried for stars located within $8\times R_{\rm search, sky}$ and $8\times R_{\rm search, PM}$ to ensure we were not artificially excluding member stars. We limited our request to stars with $\texttt{phot\_g\_mean\_mag}<18~$mag and $\texttt{astrometric\_params\_solved} = 31$ (i.e., stars having position, parallax, and PM measurements). 
We then performed a clustering analysis using the clustering algorithm \texttt{HDBSCAN} \citep{mcinnes17} 
adopting $\texttt{min\_samples} = 5$ and
\begin{equation}
    \texttt{min\_cluster\_size} = 
    \begin{cases}
        20&N_{\rm lit}<100 \\
        50&{\rm otherwise ,}
    \end{cases}
\end{equation}
with $N_{\rm lit}$ being the reported number of members by \citet{cantatgaudin_etal2020}.
The parameter \texttt{min\_samples} sets the minimum number of sources used in determining the nearest neighbor distance for each source. Hence, increasing \texttt{min\_samples} increases the mutual reachability distance among sources, and only the densest areas survive as clusters.
\texttt{min\_cluster\_size} defines the minimum number of stars required to form a cluster\footnote{We refer to the online documentation (\url{https://hdbscan.readthedocs.io/en/latest/index.html}) for further details.}.
Such parameter choice was set to maximize the recovery fraction of clusters in the \citet{cantatgaudin_etal2020} catalog. In fact, we recover almost all (1862 corresponding to $92.3\%$ of the starting sample) clusters in the catalog.
In Fig.~\ref{fig:clustering_ngc869} we present the case of NGC~869 as an example of the adopted analysis. Different panels show the distribution of field and cluster member stars in space, velocity and in the color-magnitude diagram.

For each cluster, we then selected as members those stars with membership probability $>50\%$ as provided by the clustering 
analysis performed in this work. While the definition of the membership probability treshold
is somehow arbitrary, we note here that a $50\%$ selection secures a pretty good match with the number of members 
stars provided by \citet{cantatgaudin_etal2020}. Also we tested that variations of the adopted probability treshold within 
reasonable ranges ($50\%-70\%$) do not have a significant impact on the kinematic analysis results (see Sect~\ref{sec:kin_analysis}).
For the kinematic analysis we retained only stars with reliable astrometry, i.e. sources having $\texttt{ruwe}<1.4$, $\sigma_\varpi / \varpi<0.2$, and $\texttt{astrometric\_excess\_noise}$ below the 95th percentile of all member stars (if $\texttt{astrometric\_excess\_noise\_sig}>2$, \citealt{lindegren21}). 

The cluster center was defined as the median position of member and kinematically selected stars, and the mean motion was obtained as the median of the $\mu_{\alpha}^*$ and $\mu_{\delta}$ PMs.
All velocities were corrected for perspective effects induced by the clusters' systemic motions by using the equations 
reported in \citet{vanleeuwen09} and following the approach already adopted in previous works of our group \citep{dalessandro21a,dalessandro21b,dalessandro24,dellacroce24}.
To this aim we used the catalog of mean cluster \texttt{LOS} velocities provided by \citet{tarricq21} while 
the clusters' distance was obtained as the sum of individual inverse parallax measurements weighted over their errors. 
Gaia parallaxes were corrected following prescriptions by \citet{lindegren21}.
For each cluster we checked the amplitude of the perspective corrections relative to the absolute PM measured by {\it Gaia}. We found that for only four clusters the median relative amplitude exceeds $20\%$.
The sample of recovered clusters for which it was possible to apply perspective corrections includes 1270 systems. Among them, we then selected 559 clusters with more than 100 likely member stars (after astrometric quality selection) to perform 
the kinematic analysis described in the following (Sect.~\ref{sec:kin_analysis}).

\section{Kinematic analysis}
\label{sec:kin_analysis}
\subsection{Velocity dispersion, rotation and expansion profiles}
\label{sec:vel_disp}
We analyzed the 1D kinematic properties in terms of velocity dispersion, expansion, and
rotation for all clusters in the sample by using the tangential (\texttt{TAN}) and 
radial (\texttt{RAD}) velocity projections of the the available Gaia PMs. 

To derive the velocity dispersion and rotation pattern of each selected cluster, 
we adopted a Bayesian approach based on a discrete fitting 
technique to compare simple kinematic models with individual velocities (see \citealt{cordero17,dalessandro18b,dalessandro24}).
This is a purely kinematic approach aimed at searching for relative differences among different clusters and it is not aimed 
at providing a self-consistent dynamical description.

The likelihood function for the velocities of individual stars depends on the assumptions about the formal descriptions of the rotation and velocity dispersion radial variations.
For the velocity dispersion profile, we assumed the functional form of the Plummer model \citep{plummer11}, which is simply defined by its central velocity dispersion $\sigma_0$ and its scale radius $a$:

\begin{equation}
\label{sigma2R}
\sigma^2_{\rm TAN}(R) = \frac{\sigma_{\rm TAN,0}^2}{\sqrt{1 + R^2/a^2}} \ ,
\end{equation}

where $R$ is the projected distance from the center of the cluster. 
For the rotation curve, we assumed cylindrical rotation and adopted the functional form expected for stellar systems 
undergoing violent relaxation during their early phases of evolution \citep{lyndenbell67}:

\begin{equation}
\label{rotation_pm}
\mu_{\rm TAN} = \frac{2V_{\rm peak}}{R_{\rm peak}} \frac{R}{1+(R/R_{\rm peak})^2}
\end{equation}

In Eq.~\ref{rotation_pm}, V$_{\rm peak}$ represents the maximum (in an absolute sense) of the mean motion in the \texttt{TAN} component and it corresponds to the rotational velocity at $R_{\rm peak}$.

We stress that, while young clusters may be out of equilibrium, 
not or only partially virialized, expanding and/or contracting (see for example
\citealt{kuhn19,dellacroce24,jadhav24}), 
the adoption of equilibrium models, such as Plummer and Lynden-Bell models, does not introduce any significant limitation. In fact, these models can nicely approximate flat velocity dispersion profiles, as those observed for 
expanding/contracting clusters, and solid-body like rotation, when adequately large scale factors (i.e., $a$ and $R_{\rm peak}$) are adopted. 
 
The kinematic analysis was performed by using the \texttt{emcee} (\citealt{foremanmackey13}) implementation 
of the Markov chain Monte Carlo (MCMC) sampler, which provides the posterior probability distribution function (PDF) 
for $\sigma_0$, $a$, $V_{\rm peak}$ and $R_{\rm peak}$. For the present 
analysis, we sampled the posterior PDFs by using $64$ {\rm walkers} for $20000$ steps each. For each {\rm walker}, the 
first $5000$ steps were discarded to ensure convergence. We did not account for correlation between samples.
For each quantity, the 50th-, 16th- and 84th-percentile of the PDF distributions were adopted as the best-fit value and relative errors, respectively. 
We assumed a Gaussian likelihood and flat priors on each of the investigated parameters within a  reasonably large range 
of values.
In particular, we imposed that both $a$ and $R_{\rm peak}$ have to be smaller than $5\times r_{\rm max}$, where $r_{\rm max}$ represents the distance of the farthest member star in the cluster.
We also imposed $\sigma_{\rm TAN,0}<5$ mas/yr and $-5<V_{\rm peak}<5$ mas/yr.
It is important to note that in general, since the analysis is based on the conditional probability of a velocity measurement, given the position of a star, 
our fitting procedure is not biased by the spatial sampling of the stars.
However, the kinematic properties are better constrained in regions that are better sampled (i.e. larger number of stars with available kinematic information).

\begin{figure}
    \centering    
    \includegraphics[width=0.5\textwidth]{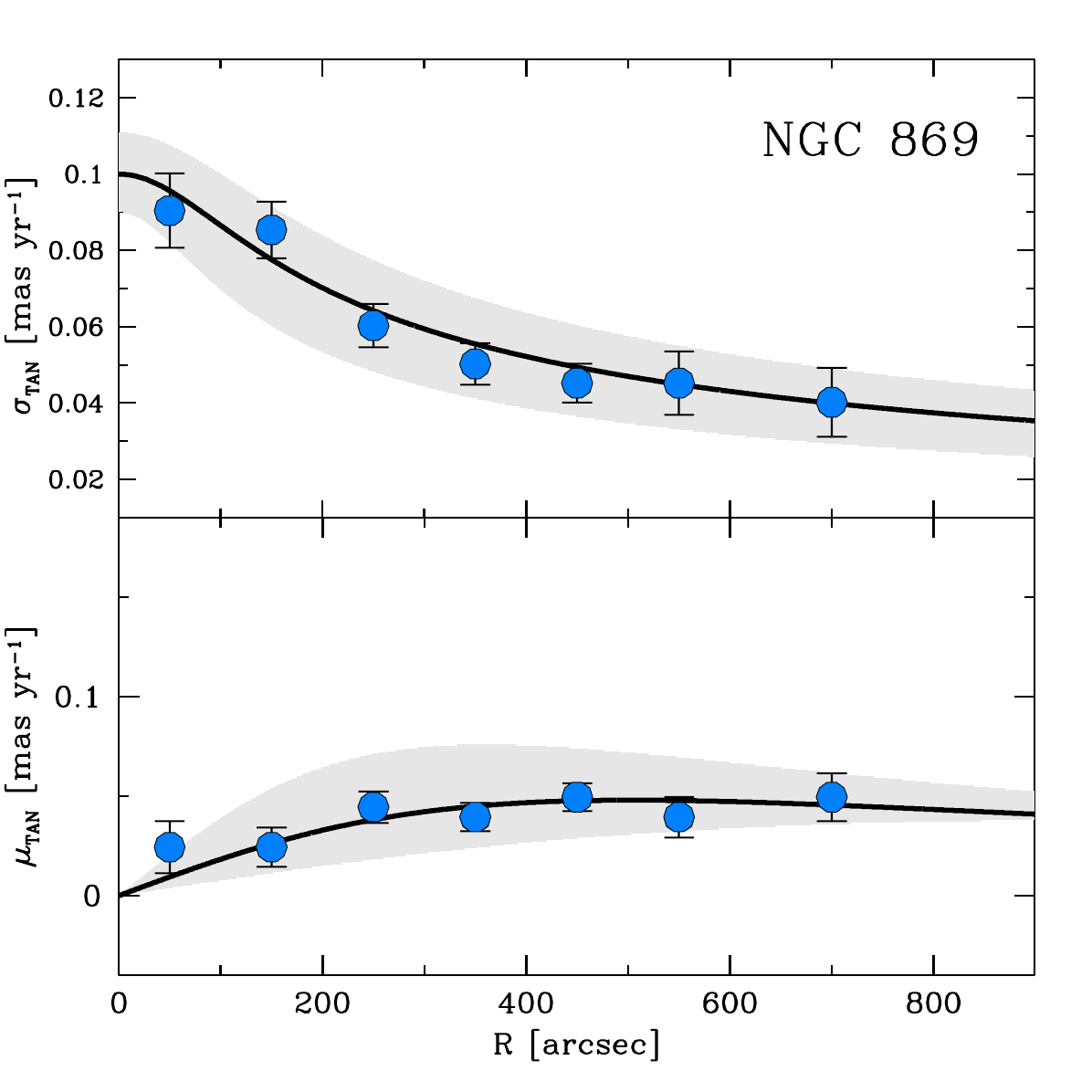}
    \caption{Velocity dispersion (upper panel) and rotation (lower panel) profiles for the OC NGC~869. 
    Blue circles represent the observed values as obtained by using a maximum-likelihood approach on binned data. 
    The black lines and shaded grey areas represent the best-fit profiles obtained from the Bayesian analysis on 
    discrete velocities described in Sect.~\ref{sec:vel_disp}}
    \label{fig:kin_869}
\end{figure}

\begin{figure}
\includegraphics[width=1\linewidth]{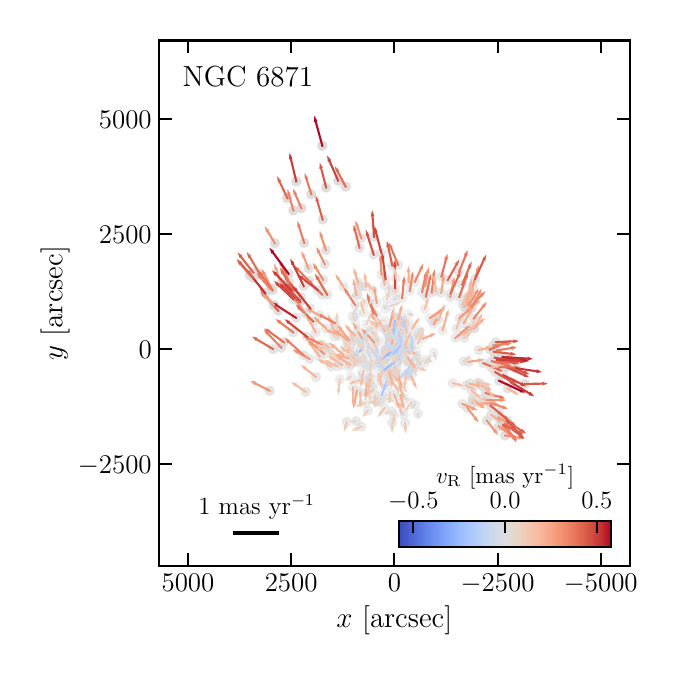}\hfill
\includegraphics[width=1\linewidth]{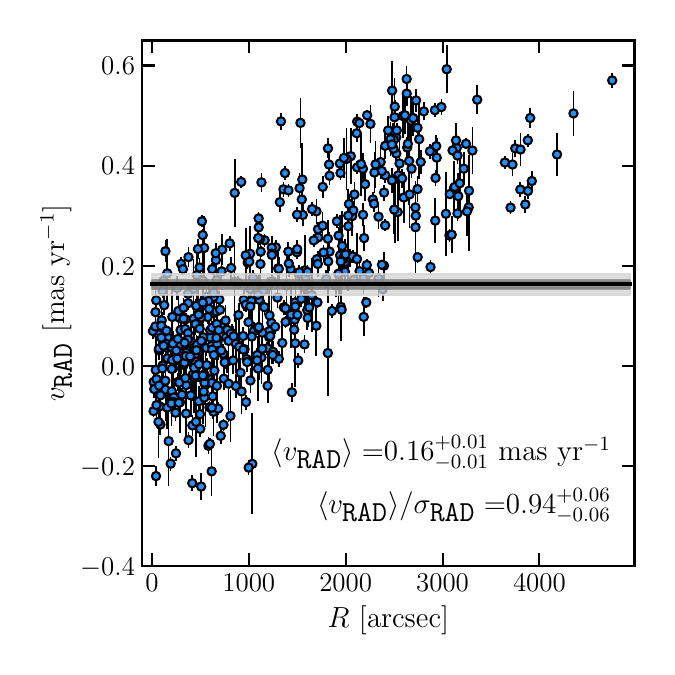}
\caption{The upper panel show the spatial distribution of members stars of NGC~6871 in Cartesian coordinates with arrows showing
the velocity vectors on the plane of the sky. The arrow lengths are proportional
to the speed on the plane of the sky while their colors map the radial
component ($v_{\rm RAD}$) of the velocity. Positive
values point outward. The bottom panel shows the distribution of members in the
$v_{\rm RAD}-R$ diagram. The black line represents the bet-fit value for $\langle v_{\rm RAD} \rangle$ and the grey dashed area its uncertainties.
}
\label{fig:expansion}
\end{figure}

As a sanity check and comparison, we also derived the rotation and velocity dispersion profiles along the \texttt{TAN} 
component by splitting the surveyed areas of each selected cluster into a set of concentric annuli, 
whose width was chosen to include at least 20 stars. 
While the number and width of the radial bins are at least partially arbitrary and their choice can potentially have an impact on the final results, this approach has the advantage of avoiding any assumption on the model description of the velocity dispersion and rotation profiles.
In each radial bin, the mean velocity and velocity dispersion and their errors were computed by following the maximum-likelihood approach described by \citet{pryor93}.
Examples of the results obtained with both the Bayesian and maximum-likelihood analyses are shown in Fig.~\ref{fig:kin_869} for the stellar cluster NGC~869.

While cluster rotation is the main focus of this paper, we characterized also cluster expansion for all clusters in our sample as we use this quantity for a comparative analysis in Sect.~\ref{sec:results}. 
To this aim we adopted the same approach described in \citet{dellacroce24}. In brief, we inferred the mean velocity
along the \texttt{RAD} component,
$<v_{\rm RAD}>$, and the velocity dispersion, $\sigma_{\rm RAD}$, in a fully Bayesian framework properly accounting for errors on individual velocities. 
Similarly to the velocity dispersion along the tangential component and rotation, we explored the parameter space by using the MCMC technique.
We assumed that the intrinsic distribution along the radial component is Gaussian with mean $\langle v_{\rm RAD} \rangle$ and velocity dispersion, $\sigma_{\rm RAD}$.
We used uniform priors within $[-10;+10]$ mas yr$^{-1}$ and $[0.001;+15]$ mas yr$^{-1}$ for $<v_{\rm RAD}>$ and $\sigma_{\rm RAD}$, respectively.
We show in Fig.~\ref{fig:expansion} the expansion pattern observed in NGC 6871, as an illustrative example. 

We verified that proper motion spatially correlated systematic errors, as reported for example by 
\citet{lindegren21} and \citet{fabricius21} for 
Gaia DR3 (see also \citealt{vasiliev19} for a study focussed on cluster kinematics with Gaia DR2), are negligible for the 
present analysis both in terms of the best-fit values of the involved kinematic quantities and their associated error budget.

\subsection{Full 3D analysis\label{sec:3d_analysis}}
For the sub-sample of clusters with an adequate number of line-of-sight (\texttt{LOS}) velocities 
and Gaia PMs ($>100$ - see Sect.~\ref{sec:results} for details) we performed a full 3D analysis.
Such an analysis has the advantage of overcoming the typical limitations
connected with projected effects, typically arising when \texttt{LOS} and plane-of-the-sky velocities are used independently, possibly hampering the detection of actual rotation signals.

\begin{figure}[!t]
    \centering
\includegraphics[width=0.5\textwidth]{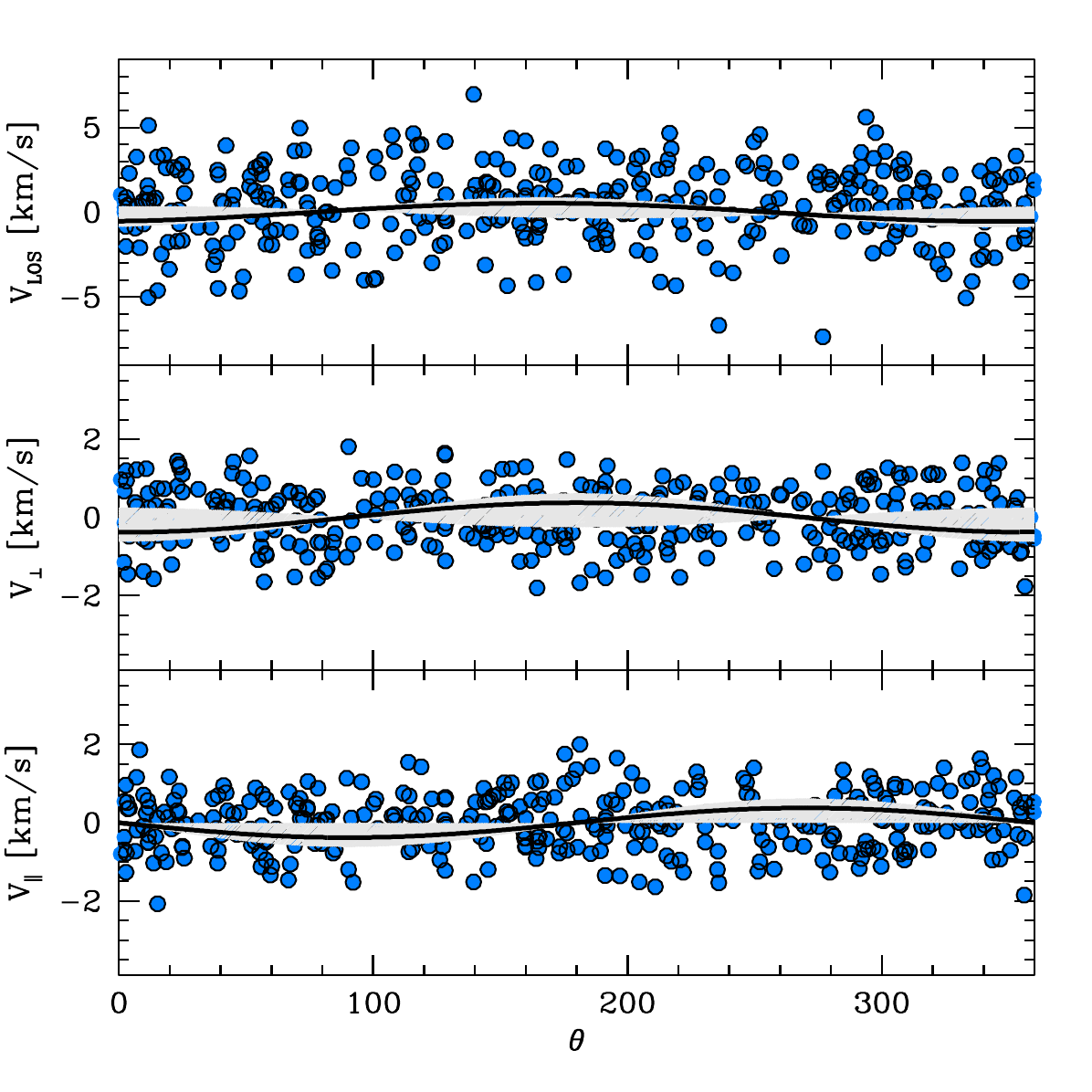}
    \caption{Distribution of the three velocity components as a function of the position angle for NGC~3532.
     The black line and the dashed grey area represent the best-fit and relative uncertanties as obtained 
    in Sect.~\ref{sec:3d_analysis}.} 
    \label{fig:kin_3D}
\end{figure}

\begin{figure}
    \centering
    \includegraphics[width=0.5\textwidth]{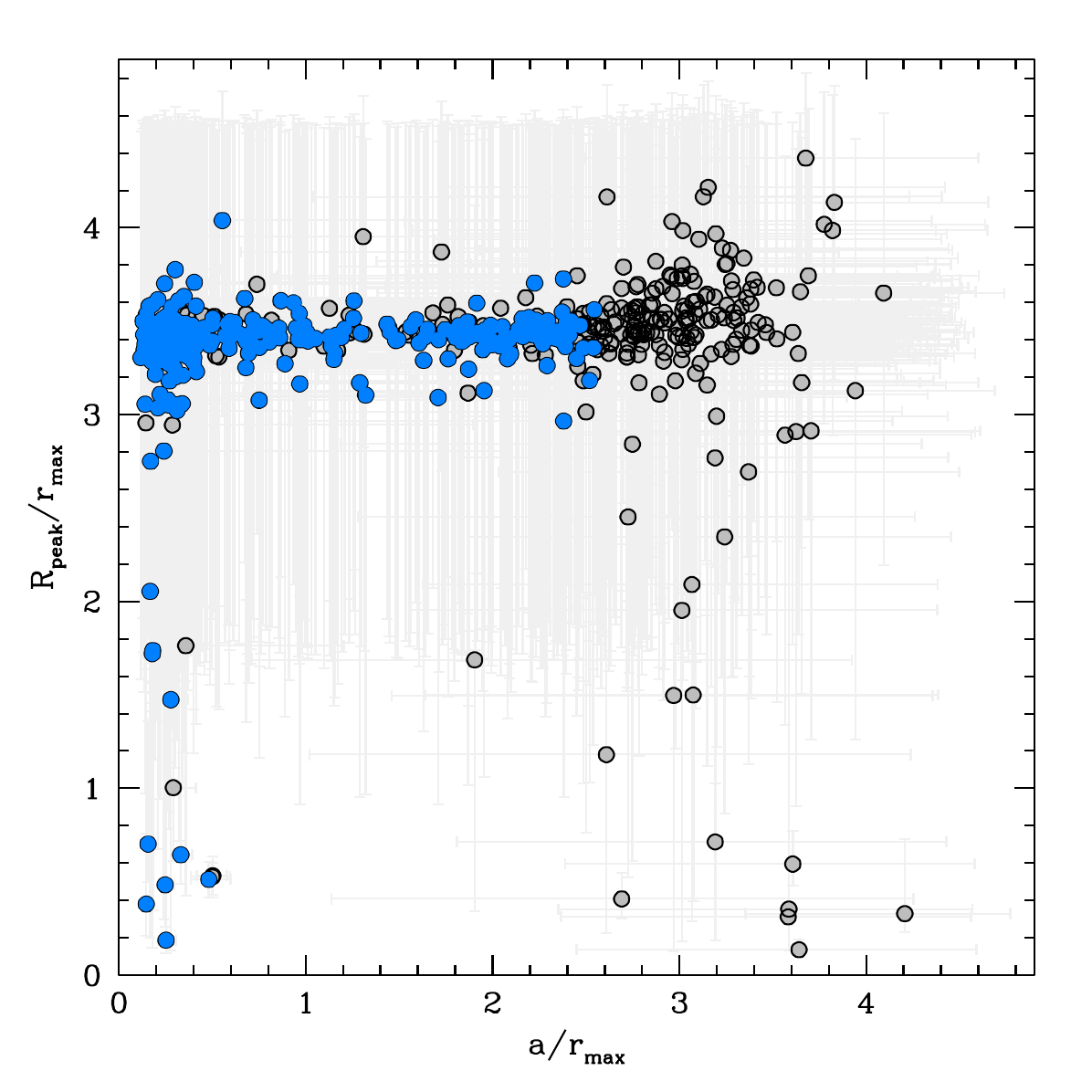}
    \caption{$a/r_{\rm max}$ and $R_{\rm peak}/r_{\rm max}$ distribution for OCs in the sample. Blue circles represent clusters selected as described in Sect.~\ref{sec:rotation}.}
    \label{fig:dist}
\end{figure}

We followed the approach described in \citet{sollima19} and \citet{dalessandro24}, which has the advantage of constraining a cluster 
full rotation pattern, estimating the inclination angle of the rotation axis ($i$) with respect to the line of sight, 
the position angle of the rotation axis ($\theta_0$) and the rotation velocity amplitude ($A$), by means of a model-independent analysis. 
We considered an average projected rotation velocity with amplitude $A_{\rm 3D}=\langle \omega R \rangle$, where $\omega$ is the angular velocity and $R$ is
the projected distance from the cluster center. Here we assume $A_{\rm 3D}$ to be independent on the distance from the cluster center. 
While of course, this represents a crude approximation of the rotation patterns expected in stellar clusters and it provides a rough 
average of the actual rotation amplitude, it is important to stress that it does not introduce any bias in the estimate of $\theta_0$ and $i$ and on their use (see Sect.~\ref{sec:prograde}).

$A_{\rm 3D}$, $i$ and $\theta_0$ were derived by solving the equations describing the rotation projection along the \texttt{LOS} ($V_{\rm LOS}$) and 
those perpendicular ($V_{\perp}$) and parallel ($V_{\parallel}$) to the rotation axes (see Eq. 2 in \citealt{sollima19}). 
While the velocity component perpendicular to the rotation axis has a dependence on stellar positions within the cluster along the line-of-sight, 
we neglected it in the present analysis as it does not affect the mean trend of the $V_{\perp}$ component, but it can only introduce an additional spread on its distribution.  
We assumed a flat distribution for $A_{\rm 3D}$ in the range $-5$ km s$^{-1}<A_{\rm 3D}<+5$ km s$^{-1}$, 
$i$ to vary in the range $0^{\circ}<i<90^{\circ}$\footnote{To secure an isotropic prior over the solid angle in the MCMC analysis, 
we adopted
a prior on sin$i$ in the range [$0,+1$].} with respect to the line-of-sight and
the position angle to be in the range $0^{\circ}<\theta_0<360^{\circ}$. $\theta_0$ grows anticlockwise from north to east and $A_{\rm 3D}$ is positive for clockwise rotation in the plane of the sky.
Following the approach already adopted for the 1D analysis, we derived the best-fit rotation amplitudes, 
position and inclination angles and relative errors by maximizing the 
likelihood function reported in Eq. 3 of \citet{sollima19} by performing a MCMC analysis by means of the 
\texttt{emcee} Python package. 
Fig.~\ref{fig:kin_3D} shows the result of the best-fit analysis along the three velocity components for the case of NGC~3532, as an example.

\begin{figure*}[!ht]
    \centering
    \includegraphics[width=1\textwidth]{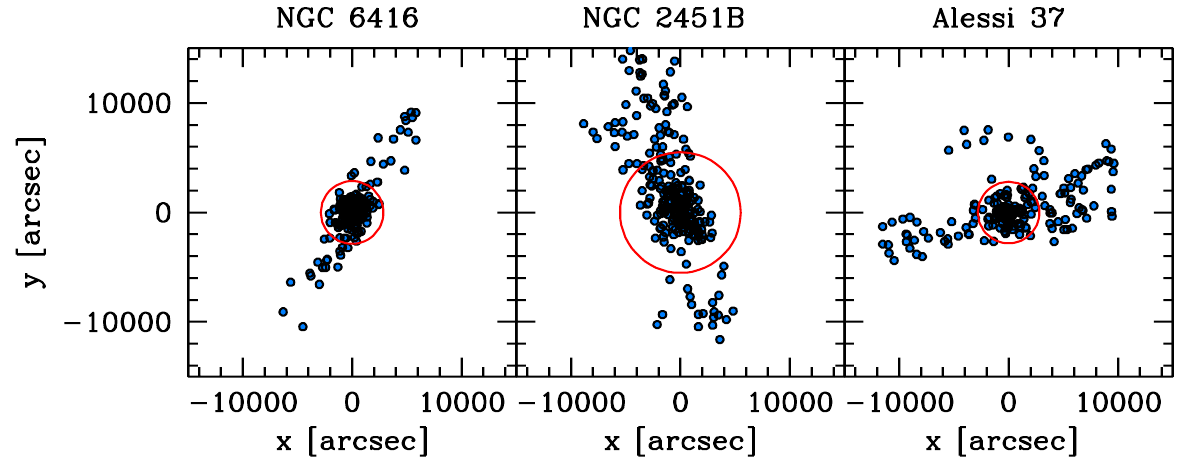}
    \caption{2D stellar distribution maps for three selected systems with clear evidence of tidal tails (see Section~\ref{sec:kinematic_characterization}).}
    \label{fig:tails}
\end{figure*}

\section{Results} 
\label{sec:results}
\subsection{Kinematic characterization}\label{sec:kinematic_characterization}
Unsurprisingly, the first interesting and general result coming out from the kinematic analysis described in Sect.~\ref{sec:kin_analysis} is that a non negligible fraction of OCs are possibly 
out of equilibrium: they are either expanding, contracting or dispersing, and they may have extra-tidal features. 
Fig.~\ref{fig:dist} shows the distribution of $a/r_{\rm max}$ versus $R_{\rm peak}/r_{\rm max}$ for clusters with more than 100 member stars. We note here that the maximum range of observed $a/r_{\rm max}$ values and the fact that $R_{\rm peak}/r_{\rm max}$ peaks around $3.5$ have not necessarily a physical meaning, but they are likely shaped by the adopted priors in the kinematic analysis (Sect.~\ref{sec:kin_analysis}).
The first criterion we used to identify candidate out-of-equilibrium clusters is based on the ratio $a/r_{\rm max}$ 
(Fig.~\ref{fig:dist}).  
Values of this ratio smaller than 1 imply that the cluster velocity dispersion profile is reasonably well fit by a 
Plummer model. On the contrary, clusters where $a$ is significantly larger than $r_{\rm max}$
would imply that the Plummer scale length parameter exceeds the extension of the cluster and the velocity 
dispersion profile is
almost flat, thus suggesting they are likely not in a dynamical equilibrium configuration. 

We observe that 182 OCs in our sample have $a/r_{\rm max}>1$ (after accounting for the uncertainties).  
For 69 clusters among them $r_{\rm max}$ significantly exceeds the value of the Jacobi radius ($r_J$; from \citealt{hunt24}), thus possibly 
suggesting they may have extra-tidal features.
At a visual inspection, we find that in some cases these clusters show a pretty spherical and symmetric configuration with just a handful of stars 
located outside $r_J$. For these systems, the definition of extra-tidal features is pretty weak and it can depend significantly on the adopted membership probability threshold. More interestingly, a fraction of OCs show extremely well defined, coherent and extended tidal tails (see \citealt{kos24,risbud25,jadhav25,sharma25} for recent results on the subject). Fig.~\ref{fig:tails} shows the 2D stellar distribution maps of a selection of these systems as an example. These tails or tidal-like structures can be related to both internal dynamical processes, interactions with the Milky Way potential and/or with the bar and spiral arms.

Fig.~\ref{fig:dist} also shows that for the vast majority of clusters $R_{\rm peak}$ is significantly larger 
than $r_{\max}$. For clusters with non null rotation ($|V_{\rm peak}|>0$) this implies that they are likely 
charaterized by solid-body rotation.

As a key information in the characterization of the dynamical state of the star clusters in the sample, we inferred their expansion or contraction state by following the approach we defined in \citet{dellacroce24}. In particular, we used the ratio between the mean radial velocity, $\mean{\vR}$, and the radial velocity dispersion, $\sigmaR$. This parameter directly evaluates the cluster expansion (i.e., $\ratioR>0$), contraction (i.e., $\ratioR<0$), or equilibrium (i.e., $\ratioR$ compatible with 0) state by using individual measurements and accounting for errors. We measured this parameter for all clusters in the sample adopting the selection presented in Sect.~\ref{sec:data}. We note that $\ratioR$, as defined, cannot probe radial trends of the expansion state and that a cluster-centric dependent analysis would be required \citep{lim20, dellacroce25}. Nonetheless, it provides a robust estimate of the cluster's internal dynamical state even for sparsely populated clusters.
We find compatible results with \citet{dellacroce24} for the overlapping age range ($<300$ Myr): expansion appears to be a dominant feature for clusters younger than $\sim30~$Myr, then the fraction of expanding clusters significantly decreases and, for older ages, the vast majority of systems appears to be compatible with a non expansion configuration.

\begin{figure*}[!ht]
    \centering
    \includegraphics[width=.49\textwidth]{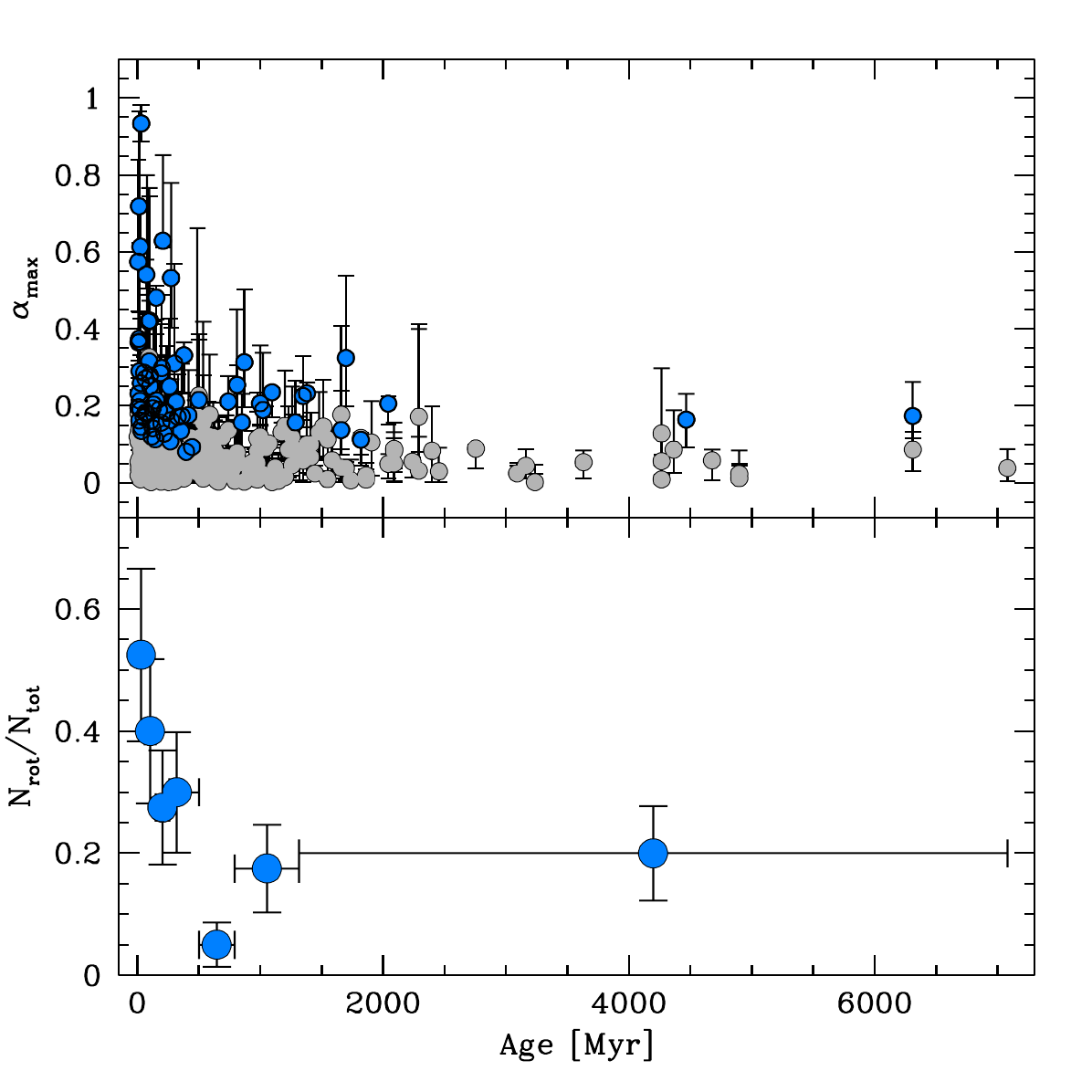}
    \includegraphics[width=.49\textwidth]{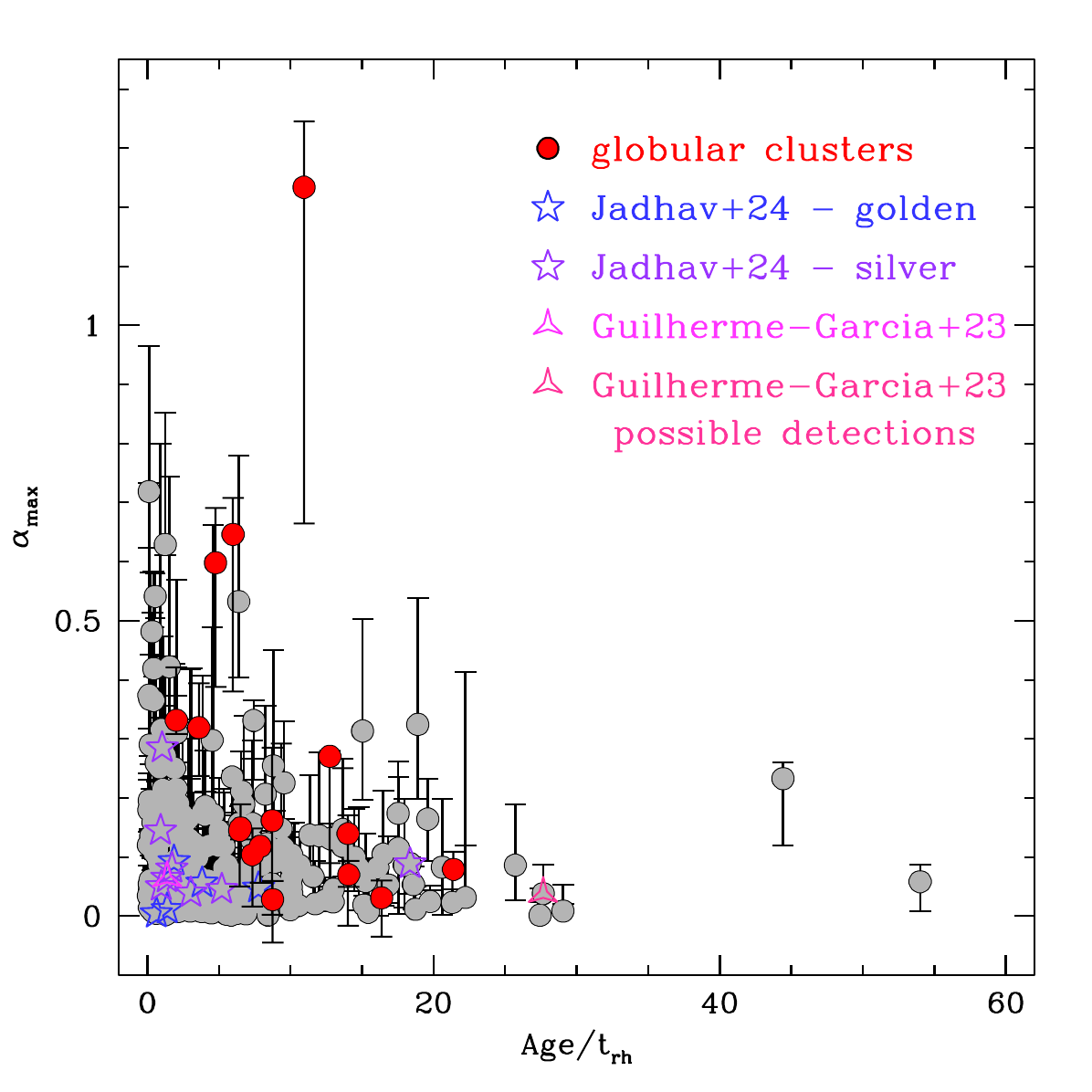} 
    \caption{{\it Left:} distribution of $\alpha_{\rm max}$ as a function of cluster age (upper panel). 
    Blue circles represent systems selected as significant rotators (Sect.~\ref{sec:rotation}). 
    The lower panel shows the fraction of significantly rotating systems N$_{\rm rot}$/N$_{\rm tot}$ as a function 
    of age. 
    {\it Right:} $\alpha_{\rm max}$ distribution as a function of clusters' dynamical age. 
    Stars and triangles are clusters identified as candidate rotators in the literature. 
    Red circles are Galactic GCs from \citet{dalessandro24}. The outlier GC (with $\alpha_{\rm max}>1$) is 
    NGC~6496}.
    \label{fig:dyn_age}
\end{figure*}

\subsection{Cluster rotation}\label{sec:rotation}
To include only clusters more likely to be in a dynamical equilibrium configuration (see Fig.~\ref{fig:dist} and Sect.~\ref{sec:kinematic_characterization}), 
for the cluster rotation analysis we selected\footnote{In addition to the selection described in Sect.~\ref{sec:data}.} only 
systems that have $(a-\sigma_a)<r_{\rm max}$ (where $\sigma_a$ represents the 16th percentile of the PDF distribution). While this choice is not necessary (see discussion in Sect.~\ref{sec:vel_disp}), we prefer to pick clusters possibly in an equilibrium configuration as they are likely to survive longer and hence they are more suitable targets to follow the possibile evolution of cluster rotation as function of their age. Finally, we further selected clusters having $r_{\rm max}<r_J$ to avoid stellar systems with prominent tidal tails or elongated sub-structures that can have a non negligible impact on the observed rotation patterns. The final sample counts 292 clusters, which are highlighted as blue circles in Fig.~\ref{fig:dist}.  

To obtain quantitative and homogeneous estimates of clusters' rotation patterns, to follow their evolution and eventually to compare the observational results 
with theoretical models and dynamical simulations, we used the observed distributions 
of velocities along the \texttt{TAN} component and we adopted the parameter $\alpha_{\rm max}$. 
This parameter  is a slightly modified version of the $\alpha$ parameter introduced by \citet{dalessandro24} 
in the context of characterizing the rotation of GCs and of their multiple populations. 
$\alpha$ is meant to incorporate in a meaningful way all the main relevant physical quantities at play in a 
single value and it has been proven to efficiently trace stellar cluster rotation patterns and to highlight in a robust way internal differences 
and deviations.
This parameter has the advantage of providing a robust measure of the relative strength of the rotation signal over the disordered motion at any radial 
range in a cluster without making any assumptions about the underlying star or mass distribution. 
By construction $\alpha$ depends on the considered cluster-centric distance and therefore a meaningful cluster-to-cluster comparison 
requires the parameter to be measured over equivalent radial portions in every system. 
We decided to measure $\alpha$ within $r_{\rm max}$ for all systems to avoid the possible impact on our ability
to detect rotating clusters due to the observed divergence
of the best-fit values of $a$ and $R_{\rm peak}$.
We verified also that the adoption of different radial selections does not have a significant 
impact on the overall relative distribution of $\alpha$ values.
$\alpha_{\rm max}$ is therefore defined as follows:

\begin{equation}
\alpha_{\rm max}=\int_0^{1} \mu_{\rm TAN}(R_{\rm l})/\sigma_{\rm TAN}(R_{\rm l}) dR_{\rm l}
\end{equation}

\begin{figure*}[!th]
    \centering
    \includegraphics[width=.49\textwidth]{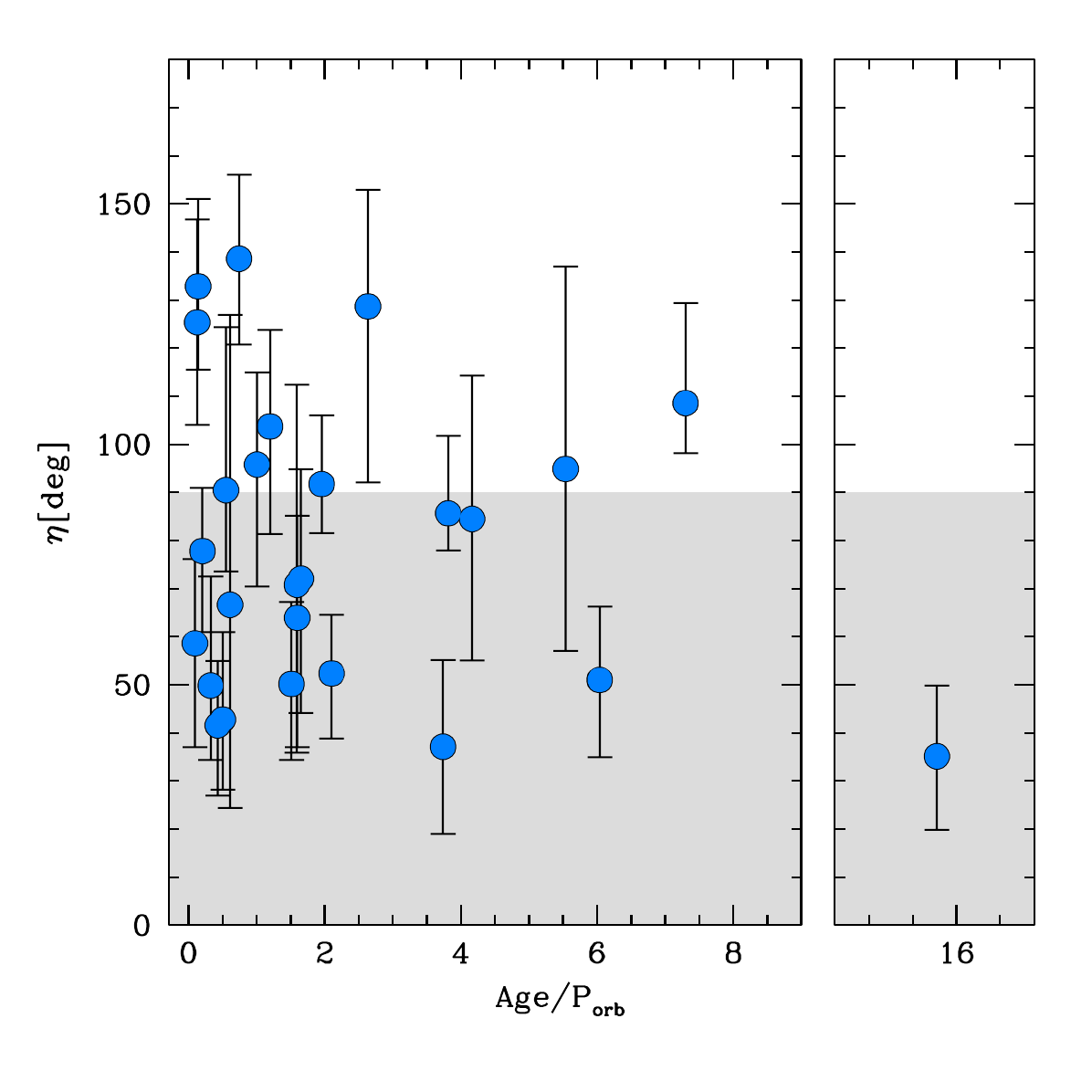}
    \includegraphics[width=.49\textwidth]{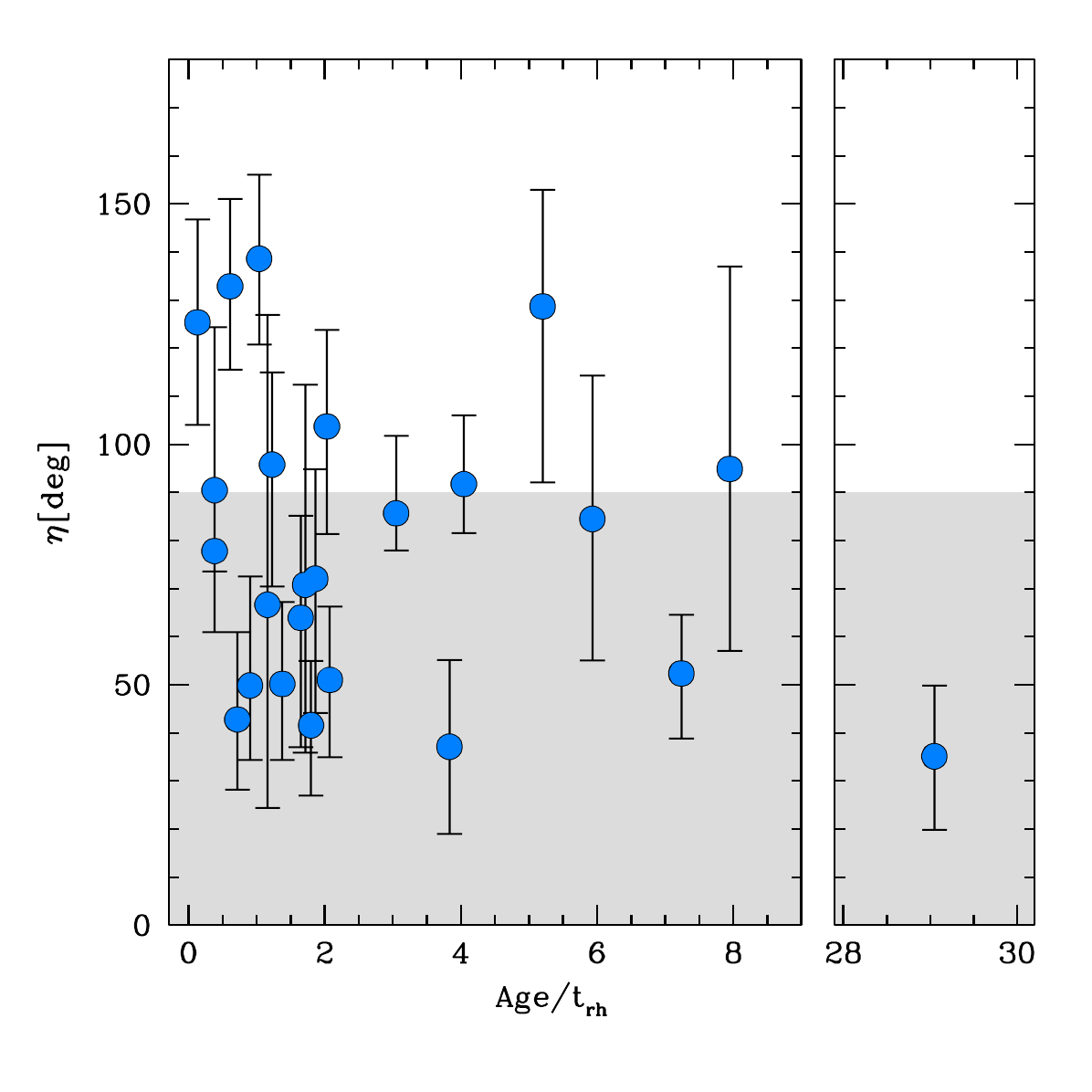} 
    \\ \vspace{-.425cm}
    \caption{Present-day inclination between the orbital angular momentum and the internal spin vector 
    (see Eq.~\ref{eq:spin_angularmomentum_angle} as a function of the number of orbits around the Galaxy. 
    Cluster ages are from \citet{cantatgaudin_etal2020} while orbital periods were computed using 
    the axisymmetric potential by \citep{bovy15}. The grey area highlights the range of $\eta$ corresponding to a 
    prograde rotation. }
    \label{fig:eta_norbits}
\end{figure*}

where $R_{\rm l}$ is the cluster-centric distance normalized to $r_{\rm max}$.
Errors on $\alpha_{\rm max}$ were obtained by propagating the posterior probability distributions obtained from the MCMC analysis for the best-fit 
rotation and velocity dispersion profiles derivation (see Sect.~\ref{sec:kin_analysis}).

The general approach of our analysis is not to focus on the detection of specific signatures of rotation,
but rather to compare the general kinematic behaviors described by all OCs in the sample in the most effective way.  

The upper panel of Fig.~\ref{fig:dyn_age} shows the distribution of $\alpha_{\rm max}$ as a function of age for the selected clusters. 
Younger clusters ($<500$ Myr) show a pretty wide range of $\alpha_{\rm max}$, which goes form $0$ to values close to 1 in a few cases, meaning that for these systems ordered and disordered motion are almost equally contributing to the cluster energetic budget. Then the distribution of $\alpha_{\rm max}$ progressively narrows down around values of $\sim0.05$ for the older systems ($t_{\rm age}>2$ Gyr). 
We have verified that the wide range of observed values of $\alpha_{\rm max}$ is not linked to the number of member stars used in the kinematic analysis thus reinforcing the idea that differences among clusters in the sample are primarily driven by intrinsically different kinematic properties.
We find that after accounting for the uncertainties on the individual $\alpha_{\rm max}$ values, 75 out of 292 clusters ($\sim26\%$) 
show significant evidence of rotation (blue circles in Fig.~\ref{fig:dyn_age}), meaning that their
$\alpha_{\rm max}$ values are larger than 0 at more than $1\sigma$. 
Interestingly, our analysis is able to detect a fraction of rotating clusters larger by a factor of $4-5$ than 
previous studies \citep{guilherme23,jadhav24} and, more importantly, it suggests that rotation is a common property of clusters at any age. 
The lower panel of Fig.~\ref{fig:dyn_age} shows the distribution of the fraction of rotating clusters ($N_{\rm rot}/N_{\rm tot}$) as a function of age. Here $N_{\rm rot}$ represents the number of clusters selected as rotators (blue points in the upper panel) and $N_{\rm tot}$ is the total number of systems in the same age range.  
We observe that $N_{\rm rot}/N_{\rm tot}$ decreases from $\sim60\%$ for systems younger than $\sim500$ Myr to $\sim15\%$ for older clusters. The significantly larger fraction of candidate rotator clusters at young ages strongly suggests that rotation is very likely imprinted during the very early stages of stellar cluster formation and evolution.  

In the right panel of Fig.~\ref{fig:dyn_age} we show the distribution of
$\alpha_{\rm max}$ as a function of the cluster dynamical age ($Age/t_{\rm rh}$), where $Age$ comes from the compilation of \citet{cantatgaudin_etal2020} and $t_{\rm rh}$ is the half-mass relaxation time. $t_{\rm rh}$ was derived by using Eq.~11 from \citet{djorg93}. We adopted as cluster mass the values reported by \citet{hunt23}\footnote{Clusters with no mass estimates do not appear in the right panel of Figure~\ref{fig:dyn_age}.}, as half-mass radius we used the radius including the $50\%$ of the member stars, as average stellar mass we assigned values adequate for a young stellar population born with a Kroupa IMF. Then the number of stars to include in the calculation is simply given by the ratio between the total mass of the cluster and the adopted average stellar mass.
A qualitative comparison with $N$-body models suggests that the observed $\alpha_{\rm max}$ distribution as a function of the cluster dynamical age appears to be compatible with the expectations for the long-term dynamical evolution and survival of stellar systems developing their rotation during their early evolution phases (see for example \citealt{tiongco17}). In fact, at any given age, cluster rotation is expected to be the remnant of a stronger early rotation (see e.g. 
\citealt{brunet12,mapelli17}) gradually weakened by the
effects of two-body relaxation \citep{bianchini18,kamann18,sollima19,dalessandro24}.
A number of numerical studies based on $N$-body simulations (e.g., \citealt{hong13,tiongco17})  show that
during the cluster long-term evolution, the amplitude of the rotation
decreases due to angular momentum redistribution and star escape.

As a comparison, in Fig.~\ref{fig:dyn_age} we report also the distribution of a sample of 16 old Galactic GCs. These systems were analyzed by \citet{dalessandro24} by using the same kinematic approach and observational strategy adopted in the present paper. For these GCs, we derived $\alpha_{\rm max}$ by adopting the tidal radius reported by \citet[][-- edition 2010]{harris96} as $r_{\rm max}$ and we adopted the $t_{\rm rh}$ values from the same compilation.
As expected and already reported in the literature (see for example \citealt{kamann18,sollima19,dalessandro24}), the rotation of GCs decreases as a function of dynamical age. 
Interestingly, while the sample of GCs is relatively small, Fig.~\ref{fig:dyn_age} shows that there is a pretty nice match between the distribution of $\alpha_{\rm max}$ and the dynamical age of OCs and GCs, thus possibly reinforcing the conclusion that rotation is imprinted at the early times of stellar clusters' life and that the long-term dynamical evolution of star clusters, mainly driven by two-body relaxation processes, is the main physical mechanism shaping its amplitude as a function of time.

We compared the observed values for rotation and expansion (see Sect.~\ref{sec:kinematic_characterization}) of the 
clusters in the sample. We find that there is a pretty weak correlation between these two kinematic quantities
(Spearman correlation coefficient $\sim0.06$), 
even when focusing only on the very young systems ($t_{\rm age}<30$ Myr), where cluster expansion is found to play 
a prominent role. 
We only note that expanding systems show a smaller fraction ($\sim38\%$) of rotators than non-expanding ones
in the same age range ($\sim50\%$) and that 
clusters undergoing significant expansion do not attain the larger rotation values observed in our analysis.

\begin{figure*}[!t]
    \centering
    \includegraphics[width=.49\textwidth]{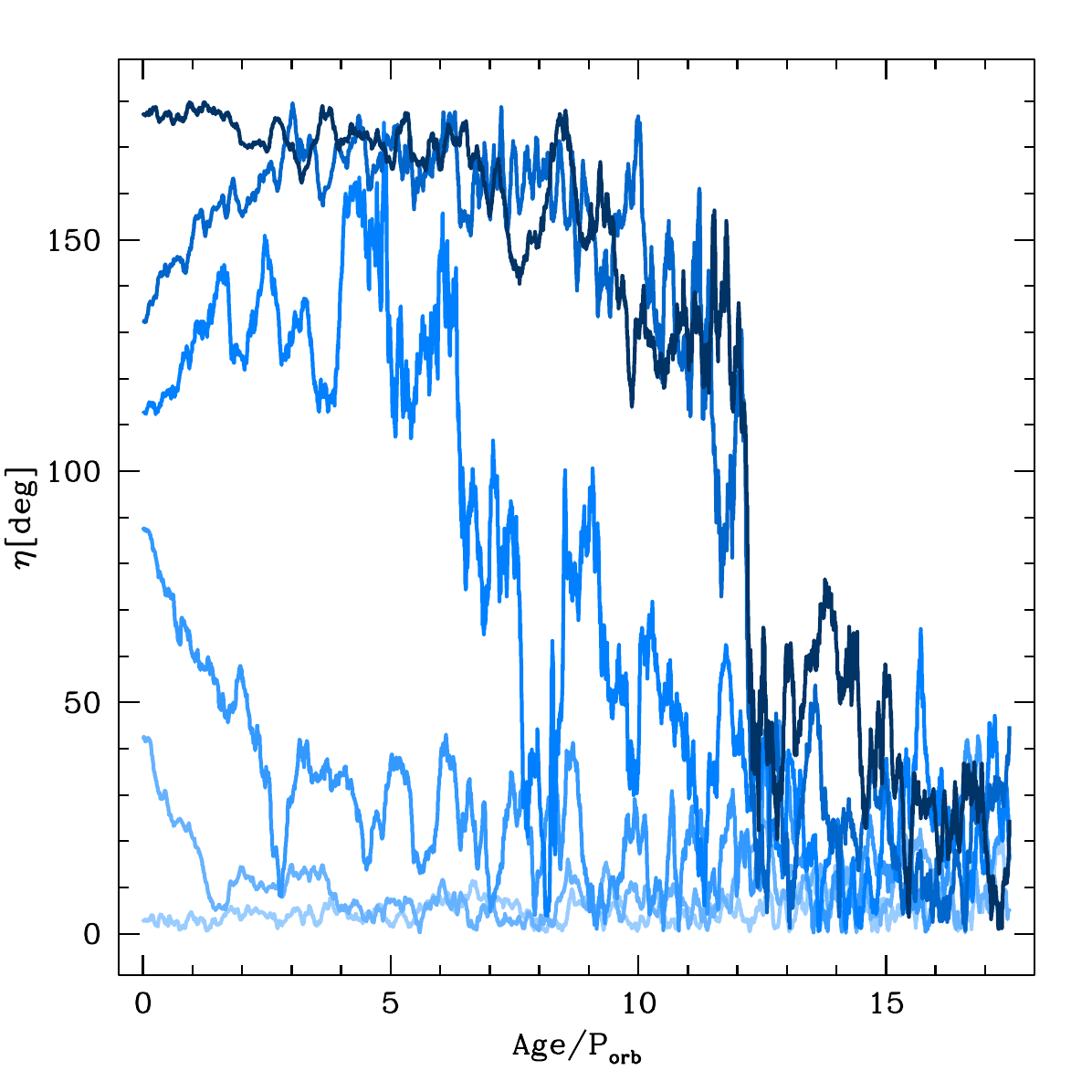}
    \includegraphics[width=.49\textwidth]{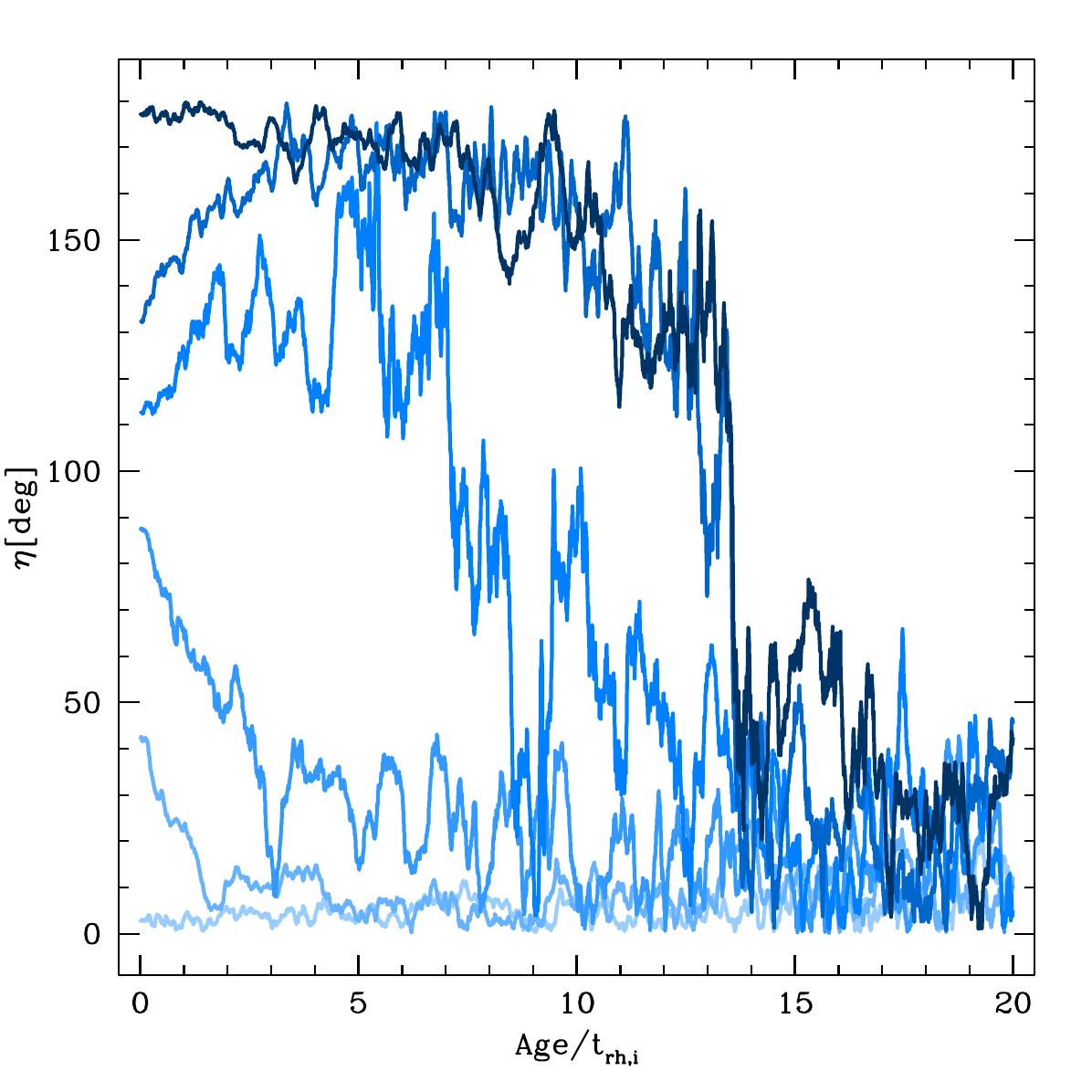} 
    \\ \vspace{-.425cm}
    \caption{Time evolution of the angle ($\eta$) between the internal angular momentum and the orbital angular momentum for the six $N$-body models described in Sect.~\ref{sec:prograde}}. 
    \label{fig:sims}
\end{figure*}

\section{Comparison with the literature}
\label{sec:lit}
Recent extensive works (e.g., \citealt{kuhn19,guilherme23,jadhav24})
based on Gaia data and available \texttt{LOS} information have provided first attempts to quantitatively and systematically study rotation among large samples of Galactic OCs.
In particular, \citet{kuhn19} studied the kinematic properties of a sample of 28 young systems 
by using Gaia DR2 PMs and found that only one cluster, namely Trumpler~15, shows evidence of rotation.
\citet{guilherme23} investigated the kinematics of 1237 clusters by using Gaia DR2 PMs and by adopting a technique aimed at reconstructing the clusters' underlying velocity field. They found that 8 clusters (and additional 9 candidates - defined as `possible' rotators; see Fig.~\ref{fig:dyn_age}) display significant rotation patterns.  
More recently, \citet{jadhav24} used a 
combination of Gaia DR3 PMs and available \texttt{LOS} for a sample 
of clusters similar to the one studied by \citet{guilherme23} finding 10 clusters with significant rotation and additional 16 candidate rotators. 
In qualitative agreement with our findings, these works would suggest that in principle young and relatively low-mass stellar clusters can be characterized by non-negligible rotation. 
However, at odds with what we observe here, they would also suggest that only a small fraction of OCs ($1\%-2\%$) shows significant signatures of rotation. 
We note that while \citet{guilherme23} and \citet{jadhav24} analyzed a similar sample of systems, because of the different adopted data and
methods, they agree only
on the rotation of three clusters, namely Stock 2, IC 2602 and Ruprecht 147.

Here we focus on a detailed comparison between these studies and the results presented in this work. A one-to-one comparison between our rotation derivations and those reported in the literature is shown in the right panel of Fig.~\ref{fig:dyn_age}. 
We find 15 clusters in common between the sample of systems analyzed in this work (Sect.~\ref{sec:kin_analysis}) and that of rotating OCs, as suggested by 
\citet{kuhn19}, \citet{guilherme23} and \citet{jadhav24}. Of them, only 4 are found to be significant rotators in our analysis based on the selection described in Sect.~\ref{sec:results}. In particular, we find NGC~2099, which is in the golden sample of \citet{jadhav24} and NGC~2547, NGC~3532 and NGC~6124, which are in their silver sample. 
We do not recover Stock~2 as significant rotator in our analysis ($\alpha_{\rm max}=0.06^{+0.04}_{-0.05}$), while IC~2602, Ruprecht~147 and Trumpler~15 are not in our selected sample. 
Along these lines, it is interesting to note that the number of candidate rotators in common between the literature and the present work increases to 32 if we do not apply any selection (neither in terms of the number of member stars nor in terms of $a/r_{\rm max}$ and $r_{\rm max}/r_J$), thus possibly suggesting that a non negligible fraction of candidate rotators in the work by \citet{guilherme23} and \citet{jadhav24} is in a non-equilibrium configuration 
and it is possibly characterized by the presence of tails and/or extended stellar structures.  
To test whether the differences between our results and those obtained by \citet{jadhav24} may arise by the different sample of cluster member stars, we performed our kinematic analysis on the same sample of selected members stars as in their works. We find that results are fully consistent with those obtained by using the selections and membership criteria described in Sections~2 and 3.  

The second important point we note is the lack in previous works of a  large fraction of rotating OCs we identify as significant rotators (see Fig.~\ref{fig:dyn_age}). While it is not easy to understand the reasons of such a difference, we argue it may be possibly related to the more efficient approach in the derivation of the kinematic properties, which is based entirely on the analysis of discrete velocities.  
We also note here that, at odds with the analysis by \citet{guilherme23} and \citet{jadhav24},
the $\alpha_{\rm max}$ parameter used in our study is based on the ratio of the rotational velocity to the velocity dispersion thus providing a measure of the relative importance of ordered and disordered motion (see also \citealt{dalessandro24} for further discussion in this respect), rather than some absolute measure of rotation. Hence, the OCs selected as significant rotators in the present study are those systems in which rotation encompasses a significant fraction of the system kinematic budget.

\section{Prograde and retrograde rotation}
\label{sec:prograde}
For the subsample of clusters for which it is possible to perform a 3D rotation analysis  (Sect.~\ref{sec:3d_analysis}) we also derived the position angle ($\theta_0$) and the inclination ($i$) of the rotation axis, providing us with all components of the internal angular momentum vector (hereafter referred to as spin, $\vec{S}$).
This allows us to calculate the relative inclination of the spin and orbital angular momentum vectors ($\vec{L}$). 

We first derive the spin vector components in the Galactocentric frame ($\vec{S}$, see Appendix~\ref{appendxix:spin_transformations} for a detailed derivation of the transformations), and we compare it with the 
orbital angular momentum vector (hereafter $\vec{L}$).
In particular, we characterize the prograde or retrograde nature of the internal rotation through the angle $\eta$ defined as
\begin{equation}
    \cos\eta = \frac{\vec{L}\cdot\vec{S}}{|\vec{L}| |\vec{S}|}\,.
    \label{eq:spin_angularmomentum_angle}
\end{equation}
Errors on $\eta$ were derived by propagating those on $\theta_0$ and $i$ in a Monte Carlo fashion and assuming the two quantities are not correlated.
From Eq.~\ref{eq:spin_angularmomentum_angle}, it follows that for $\eta<90^\circ$ the rotation is prograde (i.e., the spin and angular momentum vectors point in the same direction even if they are not aligned) and vice-versa.

The orbital angular momentum is directly derived from the clusters' mean 3D position and velocity in Galactocentric coordinates.
Moreover, we derive the cluster orbital parameters (e.g., the orbital period) within a \citet{bovy15} 
potential. All clusters in our sample have disk-like orbits, with little excursion above and below the 
Galactic plane (consistently with previous works, see e.g., \citealt{tarricq_etal22}).

In Fig.~\ref{fig:eta_norbits}, we show the angle $\eta$ as a function of the number of orbits the clusters traveled in the Galactic disk (defined
as the ratio of the cluster age to its orbital period - $Age/P_{\rm orb}$) and as a function of the dynamical age ($Age/t_{\rm rh}$). Two
main aspects could be highlighted from Fig.~\ref{fig:eta_norbits}: i) clusters that
have not yet completed a full orbit around the Galaxy uniformly
populate the range [$0^\circ$, $180^\circ$], possibly suggesting that there is
no preferential alignment of the primordial spin vector; ii) more
evolved clusters appear to have a larger fraction of prograde clusters ($\eta<90^\circ$): specifically, for clusters
with a number of orbits smaller than 1 we observe 10 systems
equally split between the prograde and retrograde configuration (left panel Fig.~\ref{fig:eta_norbits}).
For clusters that have already completed many orbits, we count 10 systems with prograde configuration ($\eta<90^\circ$)
and 6 retrograde clusters. Similarly, for dynamical ages $Age/t_{\rm rh}<1.5$ we find 11 OCs, of which 6 are prograde and 5 retrograde. More dynamically evolved systems are split between 10 in a prograde and 5 in retrograde configuration.

While the differences in the fraction between the two groups of clusters is admittedly not statistically significant and a larger sample of objects will be necessary for a more detailed statistical study of this trend, we point out that 
the evolving fraction of prograde clusters may be a hint of the effect of 
the evolution of the clusters' internal spin vector towards alignment with the orbital angular momentum.
To guide the interpretation of the results presented in Fig.~\ref{fig:eta_norbits} we have run with the 
\texttt{Petar} code \citep{wang20} a suite of six $N$-body simulations of rotating clusters with different 
initial orientations of the internal spin relative to the orbital angular momentum. 
The initial angles between the internal and the orbital angular momenta adopted in our simulations 
were equal to $0^\circ$, $45^\circ$, $90^\circ$, $115^\circ$, $135^\circ$, and $180^\circ$ thus 
spanning the full range of angles we find in our observational sample. 
Our models start with 6000 stars distributed with the density profile of a Plummer model (see e.g., 
\citealt{heggie03}) with a half-mass radius equal to 4 pc; rotation was added by using the procedure 
suggested in \citet{lynden-bell60} and reverting the tangential velocity around the rotation axis of 30 per cent of the stars in the system. Our simulations start with equal-mass models and include only the effects of internal two-body relaxation and those of the external tidal field of the host galaxy. The simulations are specifically aimed at illustrating the evolution of the angle between the internal angular momentum and the orbital one (see also \citealt{tiongco_etal18,tiongco_etal22,white25} for studies of the evolution of rotating GCs including an investigation of the orientation of the internal angular momentum).
Clusters were put on circular orbits at a distance of 8 kpc from the galactic center on the disk 
plane of the Milky Way. The potential of the Galaxy was modeled using \citet{bovy15}.
Fig.~\ref{fig:sims} depicts the resulting time evolution of the angle $\eta$ from our simulations 
and it clearly shows the gradual alignment between the internal and orbital angular momenta. 
We also note, in agreement with our observational results, that our simulations show that for the 
range of dynamical ages in our observed sample, memory of the initial distribution of prograde 
and retrograde clusters is still in part preserved. 

\section{Summary and discussion}
\label{sec:summary}
We performed a detailed kinematic analysis of a large sample of OCs, representative of the full Galactic population, with the aim of constraining their rotation and evolution with time. To this aim, we made use of Gaia DR3 PMs and \texttt{LOS} velocities obtained by large spectroscopic surveys and we adopted a Bayesian approach based on a discrete fitting technique to compare simple kinematic models with individual velocities and able to deal with relatively small statistical samples
and to account for the morphological and kinematic complexities commonly observed in young OCs. 
Following the results of a recent and similar analysis targeting GCs and their multiple populations \citep{dalessandro24}, we parametrized the strength of cluster rotation along the \texttt{TAN} velocity component by using the parameter $\alpha_{\rm max}$, which has the advantage of incorporating in an efficient and meaningful way all the main physical ingredients at play in a single value.  In addition, for a sub-sample of 26 clusters with a sufficiently large number 
of velocities along the three components, we performed a full 3D rotation analysis.

Our study shows that rotation is a common property of clusters at any age. 
In fact, about $26\%$ of the systems in our sample (corresponding to 75 clusters) are characterized by significant rotation. This result increases by a factor of about 5 the number of candidate rotator clusters with the respect to previous analysis \citep{guilherme23,jadhav24} and it finally enables an observational reading
of cluster rotation as a function of time.  
We observe that young clusters ($<500$ Myr) show a larger range of observed rotation velocities with some of them having rotation values approaching their velocity dispersions. Then, the distribution of rotation progressively narrows down at later ages. Moreover, $50\%-60\%$ of young systems are observed to be significantly rotating, while this fraction drops to $\sim15\%$ for the older ones. 
The large fraction of rotating systems as well as the large range of observed rotation values at young ($<500$ Myr) ages, suggest that rotation is very likely imprinted during the very early stages of stellar cluster formation and evolution.  In addition, while it is hard to read the distribution of $\alpha_{\rm max}$ as a function of age in terms of an evolutionary path, the general behavior observed in Fig.~\ref{fig:dyn_age} is qualitatively compatible with the expectations for the long-term dynamical evolution and survival of initially rotating clusters (e.g., \citealt{hong13,tiongco17,tiongco_etal18,kamann18,bianchini18,sollima19,dalessandro24}).
This interpretation is further strengthened by the observed distribution of $\alpha_{\rm max}$ as a function of the clusters' dynamical age and by the comparison with a sample of old GCs.

For the sub-sample of clusters with a 3D kinematic analysis, we were able to compare clusters' spin angle with their orbital one.
We observe that clusters with an orbital period larger than their age (or young dynamical ages) are equally distributed between prograde ($\eta<90^{\circ}$) and retrograde ($\eta>90^{\circ}$) values, while for systems with a larger $Age/P_{\rm orb}$ ratio (or older dynamical age) we find that they are mostly prograde. By means of a set of $N$-body simulations, we show that this trend is compatible with the expected torque-driven alignment between cluster rotation and its orbital motion \citep{tiongco_etal18,dalessandro21b,tiongco_etal22,white25}. 
Interestingly, we note that the almost homogeneous distribution of $\eta$ for young clusters would suggest that there is no preferential alignment between the original spin vector and the orbital one. We speculate here that this result can have implications on the possible physical mechanisms imprinting rotation in the emerging stellar clusters. 
First, we note that observational studies of the internal rotation of molecular clouds have found that the relative fraction of clouds with prograde or retrograde rotation  depends on the environment and vary in different galaxies: a larger fraction of internal prograde rotation has been found in the disk of M~51 and M~33 while the fraction of retrograde rotation increases for molecular clouds located in the spiral arms \citep{braine18,braine20}. In the Milky, on the other hand, observations have revealed a similar fraction of clouds with prograde and retrograde internal rotation  (see e.g., \citealt{phillips99}) 
while a recent study by \citet{liu25} found a high fraction of retrograde rotators in the spiral galaxy NGC 5064. 
Secondly, it is important to note that the hierarchical assembly of star clusters through the dynamical interactions
of smaller clumps may efficiently imprint rotation to the final cluster (e.g., \citealt{mapelli17,dalessandro21a,livernois21,dellacroce23,karam_sills_24,karam_etal25}), and alter the initial rotational kinematic of the system, resulting in significantly different configurations in terms of spin orbital vector alignment.
However, given the relatively small number of clusters and the typical uncertainties, we refrain from making any firm conclusions in this respect. 
These results should be considered rather as a proof of concept. The exquisite data provided by {\it Gaia} in synergy with large spectroscopic surveys allow us in principle to study the internal cluster rotation in three dimensions \citep{sollima19,dalessandro24} and additional efforts in this direction should be made to extend the sample for which this analysis is possible, in particular in light of the upcoming {\it Gaia} DR4 and the availability of multi object spectrographs with a large multiplexing power ($>1000$), such as WEAVE, 4MOST and MOONS.

\begin{acknowledgements}
The authors thank the anonymous referee for their careful reading of the manuscript and for their comments that 
have contributed to improve the presentation of the results. ED and GE acknowledge financial support from the INAF Data analysis Research Grant (PI E. Dalessandro) of the “Bando Astrofisica Fondamentale 2024”.
The research activities described in this paper were carried out with contribution of the Next Generation EU funds within the National Recovery and Resilience Plan (PNRR), Mission 4 - Education and Research, Component 2 - From Research to Business (M4C2), Investment Line 3.1 - Strengthening and creation of Research Infrastructures, Project IR0000034 – ``STILES - Strengthening the Italian Leadership in ELT and SKA''. 
EV acknowledge support from the John and A-Lan Reynolds Faculty Research Fund.

\end{acknowledgements}

\bibliographystyle{aa} 
\bibliography{bibliografia} 

\appendix
\section{Derivation of the internal spin vector in the Galactocentric reference frame}\label{appendxix:spin_transformations}
Let's assume we measured $PA$ (defined such that $PA=0^\circ$ points north, and increasing eastward, i.e., $PA=90^\circ$ points East) and $i$ (defined between $[0^\circ; 180^\circ]$, and pointing along the line of sight, defined as the line from the observer to the source) for a cluster centered in $(\alpha;\delta)$.
We first define the spin vector in local tangent plane coordinates
\begin{equation}
    \vec{S}_{\rm tan} = 
    \begin{pmatrix}
        \sin i ~\sin PA \\
        \sin i ~\cos PA \\
        \cos i
    \end{pmatrix}\,,
    \label{eq:spin_localtangentplane}
\end{equation}
and then derive its coordinates in the ICRS reference frame using the rotation matrix
\begin{equation}
    R_{\rm ICRS} = 
    \begin{pmatrix}
        -\sin\alpha & -\cos\alpha\sin\delta & \cos\alpha\cos\delta \\
        \cos\alpha & -\sin\alpha\sin\delta & \sin\alpha\cos\delta \\
        0&\cos\delta&\sin\delta
    \end{pmatrix}\,.
\end{equation}
Therefore,
\begin{equation}
    \vec{S}_{\rm ICRS} = R_{\rm ICRS} \cdot \vec{s}_{\rm tan}\,.
\end{equation}
Finally, we transform $\vec{s}_{\rm ICRS}$ into the Galactocentric reference frame by rotating the reference frame axes such that they coincide with those of the Galactocentric one (see, for instance, the documentation on the \texttt{Astropy} package\footnote{ \url{https://docs.astropy.org/en/latest/coordinates/galactocentric.html}.}). We did not perform any translation for the Sun position, as we were interested in the relative direction of $\vec{S}_{\rm Gal}$ with the orbital angular momentum.

\end{document}